\documentstyle[preprint,epsf,aps,prb]{revtex}
\begin{document}

\title{On the Localization of Heavy Particles in Metals.}
\author{Daniel S. Fisher and Aris L. Moustakas }
\address{Department of Physics, Harvard University, Cambridge, MA 02138}
\maketitle

\begin{abstract}
  It has been conjectured\cite{Sols,Yamada} that an impurity with
  charge $Z \ge 2$ can be localized due to its interaction with
  electrons in a metal.  The simplest case is an impurity free to move
  between only two sites, which interacts locally with \mbox{$s$-wave}
  electrons. For $Z \ge 2$ the hopping of the impurity is formally
  irrelevant and this has been argued to lead to localization. In this
  paper it is shown that other processes, in particular joint hopping
  of the impurity and one or more electrons between the sites, play an
  important role and have not been treated properly in the literature.
  Being relevant in a renormalization group sense, even when $Z \ge
  2$, these terms lead to delocalization of the impurity. Using
  bosonization, it is shown how these processes are generated from
  marginal operators that are usually neglected and the dangers of
  ignoring marginal or irrelevant operators are discussed in detail.
  Questions about implications for the more general situation of many
  sites to which the impurity can hop, are also considered.
\end{abstract}

\pacs{PACS numbers: 72.15.Rn, 73.40.Gk, 71.27.+a}

\section{Introduction}
A very interesting but subtle problem for the past quarter century has been the
low temperature (T) behavior of a heavy particle interacting with the
electronic excitations in a metal. The subtleties arise from the competition
between the tunneling of the heavy particle, which tends to delocalize it, and
the inability of the electronic degrees of freedom to adjust to the potential
of the moving particle, which tends to localize it. This difficulty is
manifested in  the orthogonality between the electronic ground states with the
impurity located at two different points in space. This phenomenon, Anderson's
orthogonality catastrophe\cite{Anderson2,Hamann}, is due to the fact that the
two ground states differ
by a very large number of very small energy particle-hole excitations. This
infinity of excitations is related to the fact that the impurity at two
different points in space creates Friedel oscillations in the electronic charge
density of the metal which, due to a difference in the phase of the
oscillations, differ from each other at arbitrary distances, implying
particle-hole deformations at arbitrarily low energies.

In the limit of weak interactions between the heavy particle and the electrons
the primary
effect of the coupling to the electrons is to induce a frictional force on the
otherwise
free tunneling motion of the particle --- although the dynamic properties of
the impurity at
low temperatures are only partially understood\cite{Leggett}. In the opposite
limit of strong
interactions, it has been argued that the particle will be strictly localized.
More
specifically, it has been claimed that a charged particle with charge $Ze \ge
2e$
tunneling between symmetric positions that are well separated spatially  will
localize
around one of the sites due to the interaction of the particle with
$s$-wave electrons.

In this paper we will argue that a heavy particle interacting with the
electrons via a small number of
channels (less than or equal to four)  {\em cannot} be localized by
the interaction because of subtle effects that have not been treated fully in
previous
work\cite{Sols,Yamada,Zawadowski,Chang}. We will primarily focus on the
simplest case in which the particle can tunnel between
only two sites that are related by symmetry. This two site problem is  related
to the Kondo
problem although there are important differences which have often been ignored
in the
literature. We primarily assume local, screened, $s$-wave, spin
independent interactions between the particle and the electronic degrees of
freedom. Due to the spin
independence of the interaction we can then neglect any possible spin of the
heavy particle and
treat the opposite spin electron species independently.

At the end of the paper we will
discuss the generalization of the problem due to the presence of three or more
sites to
which the particle can hop, and also the potential relevance of more angular
momentum
channels. We will argue that our results suggest the correct behavior for the
extended system of a particle in a periodic metal, and also have implications
for the
sharpness of X-ray edge singularities in systems with mobile deep holes and
other related
problems. The purpose of this paper is partially pedagogical, thus we work
through some
parts in substantial detail, in particular pointing out the dangers that lurk
within many
of the standard tricks, especially bosonization.

\subsection{Outline}
In the remainder of this Introduction we motivate the form of the
Hamiltonian with which we will
primarily work and explain qualitatively the effects of the orthogonality
catastrophe on
the motion of the particle as well as the effects that make it difficult to
localize.
The rest of the paper is organized as follows: In the next section (II) we
motivate
and introduce the standard and very useful method to perform the
calculations, i.e. bosonization.  The
model is introduced in the usual fermion representation of electrons
which is then
mapped into bosons. In section III  a path integral representation of the
partition function is formulated and brought into a Coulomb gas representation.
Renormalization group  flow equations are derived and
analyzed. In this way the results that were  discussed qualitatively in
the Introduction are put on a firm footing. Subsequently in Section IV, the
results, as well as possible generalizations and complications, are discussed.
Finally
in Appendix A the Coulomb gas representation of the partition function is
rederived
from the original fermion representation and in Appendix B the two site problem
is analyzed in
the absence of any symmetries other than the equivalence of the two sites.

\subsection{Physical Picture}
The Hamiltonian of the impurity--electron system has three important terms: the
non-interacting electron part
($H_o$), the hopping of the heavy particle between the two sites and the
interaction term
$U$. Thus:
\nopagebreak
\begin{equation}
H=H_o+\Delta_o(d_1^+d_2 + d_2^+ d_1) + U
\label{H}
\end{equation}
where $d_1^+, d_2^+$ are the creation operators of the impurity at sites 1, 2
and
$\Delta_o$ is the bare hopping matrix element of the impurity between
the sites. We will generally neglect any
asymmetry between the two sites. The interaction $U$ will involve terms of the
form
$d_1^+d_1 c_n^+c_{n'}$ and  $d_2^+d_2 c_n^+c_{n'}$ where $c_n^+$ are the
creation operators
of the electronic degrees of freedom. However, due to the assumed local nature
of the
potential we can rediagonalize the degrees of freedom of the electrons and be
left with
only two electronic degrees of freedom for each energy that are just composed
of those
 wavefunctions which do not vanish at the two sites. Thus the potential $U$ can
be put in a
form in which it involves two electronic states --- albeit {\em not} free
electron
eigenstates --- while all others decouple from the impurity. We can thus write
the most
general form for $U$ in the following symmetric way:
\begin{eqnarray}
U=(d_1^+d_1 + d_2^+d_2)[ V_1(c_1^+c_1 + c_2^+c_2) + V_2(c_1^+c_2 +
c_2^+c_1)] \nonumber \\ +(d_1^+d_1-d_2^+d_2)[ V_3(c_1^+c_1-c_2^+c_2) +
iV_4(c_1^+c_2-c_2^+c_1)]
\label{U1}
\end{eqnarray}
with matrix elements $V_i$ to be determined. The interchange symmetry
$1\leftrightarrow2$
is manifest in $U$. We have picked a basis for the electrons $c_i$ , $i=1,2$ so
that in the
limit that $R$, the distance between the two sites, tends to infinity, the
$c_i$'s tend to
the local $s$-wave annihilation operators at the two sites. As a result we
expect $V_2 \, , \!
V_4 \rightarrow 0$ and $V_3 \rightarrow V_1$ as $R\rightarrow \infty$. This
will be seen explicitly later.
Because there is only one heavy particle it is convenient to express
$U$ in form of Eq.(\ref{U1}) since $d_1^+d_1+d_2^+d_2=1$. Therefore only $V_3$
and $V_4$ couple the impurity
to the electrons. One can see that the four terms in Eq.(\ref{U1}) are the only
ones possible, due to the $1\leftrightarrow2$ interchange symmetry of $U$. From
these four
terms, by choosing the basis $c_i$ appropriately, one can make one term vanish
since there
are many ways we can choose normalized states  that all tend to the local
wavefunctions of
the two sites as $R\rightarrow \infty$.

The freedom of choice of states is related to a gauge symmetry. If the
system is time reversal invariant, to which we primarily restrict
consideration, then a gauge can be chosen to make the Hamiltonian real
and hence eliminate $V_4$. Note that, more generally, even in the
absence of time reversal invariance $V_4$ could be eliminated formally
at this point. But other operators would appear in the more detailed
analysis which cannot be eliminated. Although we will not analyze
these operators in detail, we will argue why they will not affect our
main results.

Thus $H$  can be written as:
\begin{eqnarray}
H=H_o+\Delta_o(d_1^+d_2 + d_2^+ d_1) + V_1(c_1^+c_1 + c_2^+c_2) \nonumber \\
+ V_2(c_1^+c_2 + c_2^+c_1) + V_3(d_1^+d_1-d_2^+d_2)(c_1^+c_1-c_2^+c_2)
\label{HU2}
\end{eqnarray}
using $d_1^+d_1+d_2^+d_2=1$.
In the next section we will explicitly derive this form and evaluate the
$V_i$'s. In the
standard manner, one can treat the relevant electronic degrees of freedom that
comprise
$c_1$ and $c_2$ as essentially {\em one dimensional} with the magnitude of {\bf
k} playing
the role of a one dimensional wavevector.

Let us now try to understand the effects of the potential on the motion of the
particle.
To start, we consider the simple limit with $R$ large so that the $V_2$ term
which couples
the two channels vanishes and $V_1=V_3$. Thus we are left with two {\em
independent}
channels, $c_1$ and $c_2$, that interact locally with the impurity. Channel 1
electrons interact with the
impurity when it is on site 1, ($d_1^+d_1=1$) with interaction strength $V_1$
and do not
interact when the impurity is on site 2, ($d_2^+d_2=1$). The opposite holds for
channel 2.
The important physics arises from the effect of the impurity at a given site on
the
electrons. If, for example, the potential is attractive, then the impurity
tends to attract
electrons towards it in order to screen its presence. Thus when the impurity is
on
site 1, it will tend to shift electrons of channel 1 towards site 1. A rather
naive picture
of this shifting is the induction of charge in a metal close to a positively
charged
object in order to screen the electric field in the bulk of the metal.
The induced charge comes
from the outer boundaries of the metal and thus from arbitrarily far
away.

A better interpretation is in terms of the wavefunctions of the electrons.
Due to the existence of the attractive potential the wavefunctions far away
from the potential center
look just like the non-interacting ones except for a phase shift. This implies
that some
extra charge density has moved in from far away to screen the impurity. Indeed
Friedel's
sum rule relates the phase shifts $\delta_\ell$ at the Fermi level to the
charge, $Ze$, that
is needed to screen a charged impurity,
\begin{equation}
Z=2 \sum_{\ell}(2\ell+1)\  {\delta_\ell}(k_F)/\pi
\end{equation}
 where the sum is over the angular momenta channels $\ell$; the sum
 over the spins yields
the factor of 2. Thus we can interpret
\begin{equation}
n_\ell=\delta_\ell /\pi
\end{equation}
as the number of electrons
per channel that need to be shifted close to the given site in order to screen
the potential. For
the present discussion we will assume that only one angular momentum channel
$\ell=0$
plays a role.

Now since the impurity can move from site to site, in order to understand the
effect of
the interaction on the dynamics of the impurity we need to know the time
dependent
amplitude of a process in which the impurity hops away from a given site for a
certain
amount of time t before it hops back to the previous site. When t is long
enough we can
view this process in the following simple way: until time $t=0$ the impurity
has been at,
say, site 1. At $t=0$ the particle tunnels to site 2 where it remains for a
time t before
tunnelling back to 1. When the particle hops away from site 1, there are
$n=\delta/\pi$
extra electrons of each spin within a screening distance from site 1 that will
move away
as the system evolves to its new ground state. Similarly there are $n$ extra
holes of each
spin near site 2. With the sites far apart, the evolution of the $s$-wave
electrons around
each site are essentially independent.

Following Schotte and Schotte\cite{Schotte} we can get a semi-quantitative
understanding of the
amplitude of the hopping process by considering $\delta/\pi =n$ with $n$ an
integer.
Between time zero and t, the $n$ extra $s$-wave spin-up electrons in channel 1
propagate as
in absence of the potential that earlier kept them near site 1. To estimate the
amplitude of the
total process, we need to find the matrix element between this evolved state at
time t and the
ground state with the particle back at site 1, i.e.\ the initial state at time
zero. Since the
radial distance from site 1 of the $s$-wave electrons can be treated as
essentially a one
dimensional coordinate, we can obtain the t-dependence of this process,
roughly, by creating the
$n$ electrons at distances a, 2a,\ldots, $n$a from site 2 with $n$a of the
order of the screening
distance. (A better approximation would involve an integral over the positions
of the extra
electrons with a weighting factor related to the wave functions in the presence
of the impurity
at site 1; but this will only modify our crude estimate by a multiplicative
prefactor.)  We thus
need to compute the amplitude\cite{Yamada,Schotte}
\begin{equation}
{A_n}(t)=\big<0\arrowvert c(a,t) \cdots c((n\!-\!1)a,t) c(na,t)c^+(na,0) \cdots
c^+(2a,0) c^+(a,0)\arrowvert0\big>
\label{A1}
\end{equation}
in a one dimensional system with no potential, with all the electrons moving at
the Fermi
velocity $v_F$ in the same direction. At long times, the antisymmetry of
Eq.(\ref{A1}) under
exchange of any two space variables fixes the form of $A_n$. When $t \gg \tau_c
= a/v_F$ , the
sum over all possible Wick pairings in Eq.(\ref{A1}) with the one dimensional
long time
propagator  \mbox{$G_o \propto [i(t-x/v_F)]^{-1}$} yields
\begin{equation}
A_n \propto \det[ t+(j-i)\tau_c]^{-1}.
\end{equation}
The determinant is of the $n \times n$ matrix with i and j subscripts. By use
of the properties
of determinants, this can be shown to yield:
\begin{equation}
A_n \propto \frac{\prod_{i<j} \left[\left(i-j\right) \tau_c \right]^2}
{\prod_{i,j} \left[\left(j-i\right)\tau_c +t \right]} \ \sim \  t^{-n^2}
\label{Asubn}
\end{equation}
for long $t \ll \tau_c$. The same result will obtain for the down spin
electrons as well as for
the $s$-wave holes around site 2.
Thus the amplitude for the full double-hop process will be $A \propto
t^{-4n^2}$. In general, with
different spin, angular momentum and site channels, $\gamma$, with $n_\gamma$
electrons moved in channel
$\gamma$, the amplitude will be $A \propto t^{- \sum_{\gamma} n_{\gamma}^2}$ .
Later we will see
that the general result for far away sites is to simply replace $n_\gamma$ by
an appropriate
phase shift $n_\gamma = \delta_\gamma / \pi$ .

It is convenient to define an exponent $\alpha_o$
\begin{equation}
\alpha_o= \frac{1}{2} \sum_{\gamma} n_\gamma^2
\label{A2a}
\end{equation}
so that the amplitude of the double-hop process will be, including dependence
on the bare hopping
amplitude $\Delta_o$ ,
\begin{equation}
A(t) \sim \Delta_o^2 \, t^{-2 \alpha_o}.
\label{A2}
\end{equation}
If the sites are not far apart, or the system is not rotationally invariant,
there will
nevertheless still be quantities analogous to $n_\gamma$, with the
interpretation as charge moved
in a ``channel'', such that  Eq.(\ref{A2}) obtains, even though the phase
shifts no longer have
any meaning, see Appendix B.

In order to understand the dynamics in the presence of the coupling to
the electrons, we make the standard argument, with the Ansatz that in
equilibrium, the heavy particle hops back and forth at a rate
$\Delta$.\cite{Leggett,Kagan} The
amplitude for this hopping can thus be guessed to be the square root
of the double-hop amplitude $A(t \sim 1/\Delta)$ since
the particle will spend time of order $1/\Delta$ at each site before hopping
back. Thus the
amplitude will be of order $A(t \sim 1/\Delta)$ for each {\em pair } of hops,
so that, from
Eq.(\ref{A2}) we have
\begin{equation}
\Delta^2 \sim \Delta_o^2 \, \Delta^{2 \alpha_o}.
\end{equation}
This has the following solution:
\begin{equation}
 \Delta = \left\{ \begin{array}{ll}
                   0 & \mbox{for $\alpha_o > 1$} \\
                   c\, \Delta_o^{\frac{1}{1-\alpha_o}} & \mbox{for $\alpha_o <
1$}
                   \end{array}
             \right.
\end{equation}
We thus see that for $\alpha_o>1$ the real hopping process will not take place
and we are thereby
lead to the conclusion that the impurity will localize on the site on which it
started undergoing
only short virtual hops back and forth to the other site. With only $s$-wave
scattering off the
impurity Friedel's sum rule yields $Z=2 \delta_o/\pi$ with $\delta_o$ the
$s$-wave channel phase
shift for the potential $V=V_1=V_3$ and the factor of 2 coming from the two
spin species. As a
result, for well separated sites,
\begin{equation}
\alpha_o =2 (\frac{\delta_o}{\pi})^2 = \frac{Z^2}{2} > 1
\end{equation}
obtaining the inequality if $Z \ge 2$, so that a charge two particle
will be localized although a charge one particle will not be.

This is the conclusion that has been reached, by this argument and more
sophisticated versions of it, by a number of
authors\cite{Sols,Yamada,Leggett,Kagan}. It seems widely
accepted - along with the extension of the result to the localization of a
particle moving on a lattice of sites. Note, furthermore that if there were
more angular momentum channels present with phase shifts of both signs, it
should be possible, by the above argument, to localize even a neutral or
charge one particle provided the phase shifts are in the regime in
which the exponent $\alpha_o > 1$. The main point of
this paper is that these conclusions are not justified. Although we will
see that it still appears to be possible to localize a particle, this {\em
cannot} be achieved by just $s$-wave scattering for {\em any} charge, and in
fact requires at least three angular momentum channels to have substantial
coupling (so that $s$ and the three $p$ channels may be sufficient).

We will
see that the approximation of neglecting the $V_2$ coupling is very
dangerous. In contrast, relaxing the approximation of $V_3=V_1$ will not change
much and the relevant phase shifts will be those associated with $V_3$. However
the
crucial $V_2$ term changes the symmetry of the problem:
  The Hamiltonian in Eq.(\ref{HU2}) with $V_2=0$ and the $c_1$ and
$c_2$ electrons uncoupled, has the continuous extra gauge symmetry
\nopagebreak
\begin{eqnarray}
\begin{array}{l}
c_1 \rightarrow c_1  \\
c_2 \rightarrow e^{i\phi} c_2
\end{array}
\label{xsym}
\end{eqnarray}
that is broken by $V_2$. The $V_2$ term mixes the two channels around the two
sites (although the mixing will be weak for large intersite
separations). This term allows processes in which one or more electrons near
one
site transform to electrons near the other. We shall see that these yield
processes in which the impurity hops from one site to the other
simultaneously with a number of electrons moving from one site to the other.
In terms of the interpretation of the exponent for the time dependence of a
process as a square of the charge transferred (such as Eq(\ref{A2a}))
we see that  the exponent for a process in which the impurity
{\em and}  a hole of each spin hop together will be
\begin{equation}
\alpha_1 = 2 (n_o-1)^2.
\end{equation}
This process is illustrated in Figure 1.
In general the process in which the impurity moves from site 1 to 2
at the same time as m holes of each spin transfer from site 1 to site 2,
will have an orthogonality exponent
\begin{equation}
\alpha_m = 2 (n_o-m)^2.
\end{equation}
Thus using the self-consistent argument for localization outlined above, we
conclude that in the case of $s$-wave scattering with $n_o=Z/2$, regardless
of the charge of the impurity, it will {\em never} become localized because
there will always be a process with $m$ pairs of holes with $m$ such that
$\left| \frac{Z}{2} -m \right| < \frac{1}{2}$ for which
\begin{equation}
\alpha_m = 2\, (n_o-m)^2 = 2 \, (\frac{Z}{2} -m)^2 < \frac{1}{2}.
\end{equation}
This process will yield a non-zero hopping rate and will delocalize the
particle.

Physically, a process in which a number of electrons hop as well as
the impurity, schematically shown in Fig.1, means that in a sense,
{\em less} of the screening cloud hops back and forth that one would
expect from the behavior of the ground state of the static impurity.
The combined process can be thought of as the tunnelling back and
forth not between the static-impurity ground states, but between
excited states, with the extra action associated with this combined process
more than compensated for by its larger matrix element (since it has a
smaller orthogonality exponent). The process with the least action
overall will dominate the impurity hopping.

We shall see that this effect can easily be missed, and indeed it seems
to have been missed in the literature\cite{Sols,Yamada,Chang}, even though a
number of authors\cite{Zawadowski,Murumatsu} have
considered ``electron assisted tunnelling'' processes in which the
heavy particle hops simultaneously with one electron (another process that
can occur). This process was also introduced in a spinless version of the
X-ray edge problem in a recent numerical work by L\'{i}bero and
Oliveira\cite{Oliveira}.
The main theoretical  difficulty is that in certain representations (e.g.\
choices of fields to bosonize) the important extra impurity-electron hopping
terms are
generated, under renormalization, from marginal terms (such as $V_2$) which
are themselves only generated from irrelevant operators. As happens all too
often, irrelevant operators cannot just be cavalierly thrown away!

\section{Model}

In this Section we introduce a simple model with short-range interactions,
show how it can be cast in the form of Eq(\ref{HU2}) and then begin to
analyze it by bosonization of the electron fields, pointing out some of the
pitfalls.

\subsection{Definitions}
We start with the Hamiltonian
\begin{equation}
H=H_o + \sum_{\sigma = \pm} U_\sigma + \Delta_o (d_1^+d_2 + d_2^+d_1)
\end{equation}
where the free electron Hamiltonian $H_o$ can be written as
\begin{equation}
H_o = \int_{\bf k} \varepsilon_{\bf k} c_{{\bf k}\sigma}^+ c_{{\bf k}\sigma}
\end{equation}
with $\int_{\bf k} = \int \frac{d^3 k}{(2\pi)^3}$ and with $c_{{\bf
k}\sigma}^+$ being the
creation operators of electrons at momentum {\bf k}, spin $\sigma$ and energy
$\varepsilon_{\bf k}$. Finally $U_\sigma$ is a short-range interaction between
electrons
and the impurity
\begin{equation}
U_\sigma = V \int_{\bf k} \int_{\bf k'} e^{-i({\bf k}-{\bf k'}) \cdot {\bf
r_1}}
c_{{\bf k}\sigma}^+ c_{{\bf k'}\sigma} d_1^+ d_1
+ V \int_{\bf k} \int_{\bf k'} e^{-i({\bf k}-{\bf k'})\cdot {\bf r_2}}
c_{{\bf k}\sigma}^+ c_{{\bf k'}\sigma} d_2^+ d_2
\label{Usigma}
\end{equation}
with V the interaction strength and ${\bf r_i}$ the position of the
i-th site. If we put the center of coordinates between the two sites
then we can set ${\bf r_1}=-\frac{\bf R}{2}$ and ${\bf r_2}=\frac{\bf
  R}{2}$ with ${\bf R}={\bf r_1}-{\bf r_2}$.

In order to eliminate the
unimportant degrees of freedom that are decoupled from the impurity we
integrate over the {\bf k}--solid angles\cite{Affleck} and are left
with a set of effectively one-dimensional degrees of freedom.
Neglecting for now the spin index $\sigma$, we define $\hat{c}_{\pm k}^+$ via:
\begin{equation}
\int \frac{dk}{2\pi} \hat{c}_{\pm k}^+ = \int \frac{d^3 k}{\left(2\pi\right)^3}
e^{\pm i
  {\bf k}\cdot \frac{\bf R}{2}} c_{\bf k}^+ = \int \frac{dk}{2\pi} \left[
\int \frac{k^2 \,d\Omega_{\bf k}}{\left(2\pi\right)^2} e^{\pm i {\bf k}\cdot
  \frac{\bf R}{2}} c_{\bf k}^+ \right]
\label{cpm}
\end{equation}
with $d\Omega_{\bf k}$ being the solid angle element in {\bf k}-space.
But now these one-dimensional Fermi operators are not properly
orthogonal. This is manifested by nonvanishing anticommutation
relations ($\{ \hat{c}_{+k}^+ , \hat{c}_{-k} \} \neq 0$). An orthonormal set of
states can be made from these that are even and odd under the interchange of
the
two sites:
\begin{eqnarray}
c_{ek}^+ = \frac{1}{\sqrt{N_e}} (\hat{c}_{+k}^+ + \hat{c}_{-k}^+)   \nonumber
\\
c_{ok}^+ = \frac{1}{\sqrt{N_o}} (\hat{c}_{+k}^+ - \hat{c}_{-k}^+)
\label{ce}
\end{eqnarray}
where the subscripts e,o denote, respectively even and odd while the
normalization constants $N_{e,o}$
\begin{equation}
N_{e,o}(k)= \frac{2k^2}{\pi} \left( 1\pm \frac{\sin{kR}}{kR} \right)
\label{Neo}
\end{equation}
are picked so that $c_{ek}$ and $c_{ok}$ satisfy one-dimensional
anticommutation relations:
\begin{equation}
\{ c_{ek}^+ , c_{ek'} \} = 2\pi \delta(k-k') \;  $etc.$
\end{equation}
{}From these states we can obtain linear combinations
\begin{eqnarray}
c_{1k}^+ = \frac{c_{ek}^+ + c_{ok}^+}{\sqrt{2}} \nonumber \\
c_{2k}^+ = \frac{c_{ek}^+ - c_{ok}^+}{\sqrt{2}}
\label{c0}
\end{eqnarray}
which transform into each other under interchange of the two sites. It
is interesting to note that $c_{1,2}^+$ are the only orthonormal
states that have this symmetry for arbitrary $kR$. To see this one
could basically define the most general pair of orthonormal states with
interchange symmetry:
\begin{eqnarray}
c_{1k}^+ = \alpha \hat{c}_{+k}^+ + \beta e^{i\theta} \hat{c}_{-k}^+  \nonumber
\\
c_{2k}^+ = \beta e^{i\theta} \hat{c}_{+k}^+ + \alpha \hat{c}_{-k}^+
\label{ctheta}
\end{eqnarray}
with $k$-dependent $\alpha, \beta, \theta$\cite{footnote1}.
Eq(\ref{ctheta}) is well defined only if $\left| \cos{\theta}
\right| > \left| \frac{\sin{kR}}{kR} \right|$. Thus for small $kR$ we
must set $\theta \approx 0$ while for large $kR$ the two sites are
decoupled and $\theta$ can take almost any value; for $\theta = 0$
Eq(\ref{ctheta})
becomes Eq(\ref{c0}) with the use of Eq(\ref{ce}). Now we can invert
Eq(\ref{ctheta}) and using Eq(\ref{cpm}), substitute into the potential
U in Eq(\ref{Usigma}). Since the only values of $k$ that play a
significant role are $k\approx k_F$ we can set $\alpha, \beta$
(equivalently $N_e$, $N_o$) to a constant evaluated at $k_F$. Thus
using
\begin{equation}
c_i^+ = \int \frac{dk}{2\pi} c_{ik}^+ \ \ \ \ \ \ \ \ \ \ \ \ \ \ \ $with$\ \
i=1,2
\end{equation}
with the integral running over $k$ in the neighborhood of $k_F$ with an
appropriate cutoff of order $k_F$, we get an expression for the
potential identical to Eq(\ref{U1}). Henceforth we will choose $\theta
=0$ for all $k$ which yields $V_4 =0$ in Eq(\ref{U1}) thereby
explicitly exhibiting the time-reversal invariance.

Furthermore we can obtain
the other coefficients $V_i$ in Eq(\ref{U1}) starting from
Eq(\ref{Usigma}) by using the relation between the free electron density of
states per
spin at $\varepsilon = \varepsilon_F$, $\rho_F$, with the Fermi
momentum $k_F$; after rescaling the Fermi velocity to be one, $2\pi^2
\rho_F=k_F^2$. We then get:
\begin{eqnarray}
\begin{array}{l}
V_1= \pi \rho_F V \\
V_2= \pi \rho_F V \frac{\sin{k_F R}}{k_F R} \\
V_3= \pi \rho_F V \sqrt{1- \left( \frac{\sin{k_F R}}{k_F R} \right)^2}
\end{array}
\label{Vi}
\end{eqnarray}
It is also instructive to write $U$ using the even-odd states from
Eq(\ref{ce}). Defining
\begin{equation}
c_{e,o}^+ = \frac{c_1^+ \pm c_2^+}{\sqrt{2}}
\end{equation}
we obtain:
\begin{equation}
U=V_1 \left( c_e^+ c_e + c_o^+ c_o \right) +
V_2 \left( c_e^+ c_e - c_o^+ c_o \right) +
V_3 \left( d_1^+ d_1 - d_2^+ d_2 \right) \left( c_e^+ c_o + c_o^+ c_e
\right)
\label{Ueo}
\end{equation}
Comparing Eq(\ref{Ueo}) with Eq(\ref{U1}) we see that (with $V_4=0$) the
non-interacting part of U is diagonal in the even-odd representation
while the interacting part is diagonal in the $c_{1,2}$
representation.

Finally, we make the standard change of variables for a two state
system, i.e.
\begin{eqnarray}
d_2^+ d_2 - d_1^+ d_1 &=& \sigma_z \nonumber \\
d_1^+ d_2 + d_2^+ d_1 &=& \sigma_x.
\end{eqnarray}
In this representation the impurity in site 1 (2) is in state $-$ ($+$) of
the $\sigma_z$ operator. Thus $d_1^+ \left| 0 \right> = \left| -
\right>$ while  $d_2^+ \left| 0 \right> = \left| +
\right>$ where $\left|0 \right>$ is the ground state of $H_o$.

Summarizing, the Hamiltonian can be written as
\begin{equation}
H=H_o + \Delta_o \sigma_x + U
\label{Heff}
\end{equation}
with
\begin{equation}
U=V_1 \left( c_1^+ c_1 + c_2^+ c_2 \right) +
V_2 \left( c_1^+ c_2 + c_2^+ c_1 \right) +
V_3 \left( c_2^+ c_2 - c_1^+ c_1 \right) \sigma_z
\end{equation}
and with $V_i$ given by Eq(\ref{Vi}) and $c_i$ given by Eq(\ref{c0})
(where $i=1,2$) while
\begin{equation}
H_o=\sum_{i=1,2} \int \frac{dk}{2\pi} \varepsilon_k c_{ik}^+ c_{ik}
\end{equation}
and
\begin{equation}
\varepsilon_k = \frac{k^2}{2m} - \frac{k_F^2}{2m} \approx \left( k-k_F
\right)
\end{equation}
with  Fermi velocity set equal to unity.

The electrons that interact with the impurity are thus effectively two
species of one-dimensional fermions moving only to the right, with
those to the left of the origin corresponding to incoming electrons
while those to the right of the origin corresponding to outgoing electrons.

At this point it is useful to pause and consider the symmetries of the
effective Hamiltonian in Eq(\ref{Heff}). There is a global $U(1)$
gauge symmetry --- of the electron phase --- and a discrete interchange
symmetry $1 \leftrightarrow 2$. However note that in the absence of the $V_2$,
there would be an {\em extra} gauge symmetry, that of Eq(\ref{xsym}).
Although in some formulations\cite{Zawadowski,Murumatsu} it appears that $V_2$
can be made to
disappear, this is potentially dangerous as $V_2$ breaks the
artificial extra gauge symmetry and the formally irrelevant operators which
break the symmetry should thus be retained.

It is instructive to see how the problem with trying to get rid of
$V_2$ can be seen in the fermion
representation; in Appendix B the analysis will be done in
considerable detail using the boson representation introduced
in the next sub-section. Using the even-odd representation of
Eq(\ref{Ueo}), one can indeed rediagonalize the even-odd channels and
absorb the $V_1$, $V_2$ terms into $H_o$. This leaves the long time Green's
functions of the even-odd channels unaffected but changes the short
time behavior (see, for example, Nozi\`{e}res and De
Dominicis\cite{NdD}). Thus
\begin{equation}
\left<c^+_e(\tau)c_e(0)\right> \neq \left<c^+_o(\tau)c_o(0)\right>
  \label{shorttimeG}
\end{equation}
for $\tau$ small, of order the cutoff $\tau_c$.
Performing perturbation theory in $V_3$ to second order  we get a
correction in $U$
\begin{equation}
\delta U \propto V_3^2 \left\{ c_e^+ c_e \left( \left< c_o c_o^+
\right> - \left< c_o^+ c_o\right> \right)+ c_o^+ c_o \left( \left< c_e c_e^+
\right> - \left< c_e^+ c_e\right> \right) \right\}.
  \label{deltaU}
\end{equation}
In general, $\left<c^+ c\right>\neq \left<c c^+\right>$ due to the
nonlinear dispersion of fermions away from $k_F$, in particular particle-hole
asymmetry. Thus, from Eq(\ref{shorttimeG}) we see that the asymmetry
between the even-odd channels reappears in perturbation theory, due to
the short time (high-energy) details. As a result we must retain the $V_2$
term in the Hamiltonian.

\subsection{Bosonization}

In order to proceed it is necessary to find a representation that
focuses on the essential low energy parts of the problem. Then, even
if the problem is not exactly solvable, one can at least hope to be
able to understand the physics and predict the low energy behavior. The most
commonly used representations are boson representations of the
pseudo-one dimensional fermions.

The basic strategy of bosonization is to try to mimic the low energy
physics of the Fermi system, which can only be done exactly for a
particularly simple system of exactly linear one-dimensional bands
with a specific form of the cutoff. In more general situations, it is
hoped (or, better, demonstrated!) that the high energy terms that are
ignored --- for example particle-hole excitations far from the Fermi
surface --- only serve to give finite renormalizations of the basic
parameters of the dominant low energy operators in the Hamiltonian.
High energy properties --- for example the fermion
anticommutation relations --- are, ipso facto, only handled approximately.
What is more important, but unfortunately sometimes forgotten, is that
terms that are formally irrelevant at low energies can, either on
their own, when combined with other terms, or under canonical
transformations, produce relevant or marginal terms that affect the
physics. As we shall see, this is the case for the present problem.

For now we will proceed in the conventional manner. Let us then start
with the noninteracting Hamiltonian, $H_o$. If we are primarily
interested in energies close to $\varepsilon_F$, we can extend the
linear dispersion relation to all energies. Thus if we set the origin
of $k$ at $k=k_F$ for convenience, we will get by Fourier
transforming:
\begin{equation}
H_o= \sum_{j=1,2} \int_{-\infty}^{\infty} dx \left[ \Psi_j^+(x)
\left(-i \frac{\partial}{\partial x} \right) \Psi_j(x) \right]
\label{Ho}
\end{equation}
where $x$ is the conjugate variable to $k$ and  $\Psi_j(x)$ is the
Fourier transform of $c_{jk}$:
\begin{equation}
\Psi_j(x) = \int_{-\infty}^{\infty} \frac{dk}{2\pi} e^{ikx} c_{jk} \ \
\ \ \ \ \ \ \ \ \ \ \ \ \ j=1,2
\end{equation}
Note that there are only {\em right} moving fermions in the system
since there is only one Fermi point, i.e. one $k$-value at which
$\epsilon_k=\epsilon_F$.

The operators at the impurity sites are given in terms of
\begin{equation}
c_j = \int \frac{dk}{2\pi} c_{jk} = \Psi_j(0) \ \ \ \ \ \ \ \ \ \ \ \
\ \ \ j=1,2
\end{equation}
which identifies $c_j$ as the $x=0$ creation operator in the one
dimensional picture. Roughly speaking, $\Psi_j(x)$ with $x<0$ corresponds to
incoming $s$-wave
electrons around site j while with $x>0$, it corresponds to outgoing $s$-wave
electrons. Thus the time reversal operator acting on $\Psi_j(x)$ will
give:
\begin{equation}
\hat{T}\Psi_j\left(x\right)=\Psi_j\left(-x\right)
  \label{TPsi}
\end{equation}
since it transforms incoming to outgoing electrons and vice-versa.

At this point we can introduce the bosonic fields $\Phi_j(x)$ by\cite{Shankar}:
\begin{equation}
\Psi_j^+(x)= \frac{1}{\sqrt{2\pi\tau_c}} e^{i\Phi_j(x)}
\label{psi}
\end{equation}
with $\tau_c^{-1} \propto k_F$ the characteristic cutoff frequency of the
order the Fermi energy, and
\begin{equation}
\Phi_j(x) = \sqrt{\pi} \left[ \phi_j(x) - \int_{-\infty}^x \Pi_j(x')
dx'\right]
\label{Phij}
\end{equation}
with $\phi_j$ and $\Pi_j$ satisfying appropriate commutation
relations:
\begin{equation}
\left[ \phi_j(x) , \Pi_i(y) \right] = i \delta_{ij} \frac{\tau_c}{\pi
  \left( \tau_c^2 + \left( x-y \right) ^2 \right) }
\end{equation}
In the continuum limit $\tau_c \rightarrow 0$, this commutation
relation approaches $i \delta_{ij} \delta(x-y)$. From Eq(\ref{TPsi})
and Eq(\ref{psi}) it can be seen that $\Phi_j(x)$ transforms under
time reversal as follows:
\begin{equation}
\hat{T} \Phi_j(x)=-\Phi_j(-x).
  \label{TPhi}
\end{equation}

Expanding $\phi_j(x)$
and $\Pi_j(x)$ in terms of their Fourier components,
\begin{eqnarray}
\phi_j(x)&=& \int_{-\infty}^{\infty}
\frac{d\epsilon}{2\pi\sqrt{2|\epsilon|}} \left[ \phi_j(\epsilon)
e^{i\epsilon x} + \phi_j^+(\epsilon) e^{-i\epsilon x} \right]
e^{-\frac{|\epsilon|\tau_c}{2}} \nonumber \\
\Pi_j(x)&=& \int_{-\infty}^{\infty}
\frac{d\epsilon |\epsilon|}{2\pi\sqrt{2|\epsilon|}} \left[-i \phi_j(\epsilon)
e^{i\epsilon x} + i \phi_j^+(\epsilon) e^{-i\epsilon x} \right]
e^{-\frac{|\epsilon|\tau_c}{2}}
\end{eqnarray}
and inserting these expressions in Eq(\ref{Phij}), $\Phi_j$ can be
written as:
\begin{equation}
\Phi_j(x)= \int_{0}^{\infty}
\frac{d\epsilon}{\sqrt{2\pi\epsilon}} \left[ \phi_j(\epsilon)
e^{i\epsilon x} + \phi_j^+(\epsilon) e^{-i\epsilon x} \right]
e^{-\frac{\epsilon\tau_c}{2}}
\label{Phij+}
\end{equation}
which involves only the positive energy parts. Subsequently inserting
Eq(\ref{Phij+}) in Eq(\ref{psi}) and then in Eq(\ref{Ho}) the non-constant
part of $H_o$ becomes
\begin{equation}
H_o=\sum_{j=1,2} \int_{0}^{\infty} \frac{d\epsilon}{2\pi} \epsilon
\phi_j^+(\epsilon) \phi_j(\epsilon) e^{-\epsilon \tau_c}
\label{Hophi}
\end{equation}
with
\begin{equation}
\left[ \phi_i( \epsilon), \phi_j^+( \epsilon') \right]= \delta_{ij}
2\pi \delta( \epsilon-\epsilon')
\end{equation}
Finally using the standard expression\cite{Shankar}
\begin{equation}
\Psi_j^+(x) \Psi_j(x) = \frac{1}{2\pi}
\frac{\partial\Phi_j(x)}{\partial x},
\end{equation}
the potential U is found to be:
\begin{equation}
  U=\frac{V_1}{2\pi} \left[ \frac{\partial \Phi_1(0)}{\partial
    x}+\frac{\partial \Phi_2(0)}{\partial x}\right]
+ \frac{V_2}{\pi \tau_c} \cos[\Phi_1(0)+\Phi_2(0)] +
 \frac{V_3}{2\pi} \sigma_z \left[ \frac{\partial\Phi_2(0)}{\partial x}
 - \frac{\partial \Phi_1(0)}{\partial x} \right].
\end{equation}
The form of U may be simplified by introducing Bose fields
corresponding to collective modes for excitations that are even, $\Phi_e(x)$,
and odd,
$\Phi_o(x)$, about the center of symmetry of the pair of sites:
\begin{eqnarray}
  \Phi_e=\frac{1}{\sqrt{2}} \left( \Phi_1 + \Phi_2 \right) \nonumber \\
 \Phi_o=\frac{1}{\sqrt{2}} \left( \Phi_2 - \Phi_1 \right)
\label{phieo}
\end{eqnarray}
The $\phi_{e,o}$ and $\Pi_{e,o}$ can be defined equivalently. In terms
of the new variables $H_o$ remains in diagonal quadratic form (i.e.
the indices $j=1,2$ in Eq(\ref{Hophi}) are replaced by $j'=e,o$),
corresponding to free bosons and U becomes
\begin{equation}
  U=\frac{V_1}{\sqrt{2}\pi} \frac{\partial \Phi_e(0)}{\partial x}
+ \frac{V_2}{\pi \tau_c} \cos[\sqrt{2} \Phi_o(0)] +
 \frac{V_3}{\sqrt{2}\pi} \sigma_z \frac{\partial\Phi_o(0)}{\partial x}
\label{Ueo'}
\end{equation}

The symmetries of the problem are manifest in Eq(\ref{Ueo'}), $\Phi_e
\rightarrow \Phi_e + c$ corresponding to the global gauge invariance,
$\sigma_z \rightarrow -\sigma_z$ with  $\Phi_o \rightarrow -\Phi_o$
corresponding to the interchange symmetry, and  $\Phi_o
\rightarrow \Phi_o + \pi\sqrt{2}$  corresponding to $c_j^+ \rightarrow
- c_j^+$ for $j=1, 2$.
Note that the even mode is completely decoupled from the
impurity in Eq(\ref{Ueo'}) and will therefore not play a
role. Formally it can be eliminated by a unitary transformation
 \mbox{$H \rightarrow \Lambda_e H \Lambda_e^{-1}$} with $\Lambda_e =
 \exp[-i\frac{V_1}{\sqrt{2}\pi}\Phi_e(0)]$; we perform this
 transformation and henceforth only consider the potential
 Eq(\ref{Ueo'}) without the $V_1$ term.

In addition, Eq(\ref{Ueo'})
 also has the symmetry:  $\Phi_{e,o}
\rightarrow -\Phi_{e,o}$ and $x \rightarrow -x$ corresponding to
time reversal  invariance from Eq(\ref{TPhi}). If the system were {\em not}
time reversal invariant, then one could have $V_4 \neq 0$.
 Indeed, the lowest order time reversal
 symmetry breaking term is $\sigma_z
\sin\left[\sqrt{2}\Phi_o(0)\right]$ which is exactly the $V_4$ term.
  However,  we argued (above Eq(\ref{HU2})) that $V_4$ can always be
  chosen to be zero. But in the absence of time reversal invariance,
 such terms as $\frac{\partial\Phi_o(0)}{\partial x}
 \sin\left[\sqrt{2}\Phi_o(0)\right]$ can also appear, essentially from
 nonlinear dispersion of the fermions away from $k_F$ and energy
 dependence of the scattering, that breaks the time reversal symmetry of
 the even or odd channels. Although
 these appear to be irrelevant, they cannot simply be eliminated  because they
  generate a  $\sigma_z \sin\left[\sqrt{2}\Phi_o(0)\right]$
 term after the unitary transformation of Eq(\ref{Lambda}) is
 performed. In order to eliminate such terms one has to pick a gauge
 or, equivalently a basis
 for the fermions (i.e. pick appropriate $\alpha$, $\beta$, $\theta$
 in Eq(\ref{ctheta})) which creates a $V_4$-term that exactly cancels
 the generated  $\sigma_z \sin\left[\sqrt{2}\Phi_o(0)\right]$ term. In
 effect, one would thus obtain a set of almost time-reversal invariant
 low energy excitations, and our main results would still obtain.

But danger lurks: even with full time reversal invariance similar
 terms to those discussed above will invalidate a related
 form of bosonization that we now discuss. It is tempting to find a
 way to get rid of the $V_2$ term by a different choice of bosonization.
 One way to do this is to start with U in the form of Eq(\ref{Ueo})
 and bosonize the fields $c_{ek}$ and $c_{ok}$. In this case one has to
introduce the fields $\Psi'_e(x)$ and $\Psi'_o(x)$ in an analogous way
to $\Psi_1(x)$ and $\Psi_2(x)$ in Eq(\ref{psi}):
\begin{eqnarray}
\Psi'_e(x)=\frac{1}{\sqrt{2\pi\tau_c}} e^{-i\Phi'_e(x)} \nonumber \\
\Psi'_o(x)=\frac{1}{\sqrt{2\pi\tau_c}} e^{-i\Phi'_o(x)}
\end{eqnarray}
with $\Phi'_{e,o}(x)$ different from  $\Phi_{e,o}(x)$ in
Eq(\ref{phieo}). Subsequently by introducing linear combinations of
these fields
\begin{eqnarray}
 \Phi_a&=&\frac{1}{\sqrt{2}} \left( \Phi'_e + \Phi'_o \right) \nonumber \\
 \Phi_b&=&\frac{1}{\sqrt{2}} \left( \Phi'_e - \Phi'_o \right),
\label{phiab}
\end{eqnarray}
U would take the form:
\begin{equation}
  U=\frac{V_1}{\sqrt{2}\pi} \frac{\partial \Phi_a(0)}{\partial x} +
 \frac{V_2}{\sqrt{2}\pi} \frac{\partial\Phi_b(0)}{\partial x}
+ \frac{V_3}{\pi \tau_c} \sigma_z \cos[\sqrt{2} \Phi_b(0)]
\label{Uab}
\end{equation}
with gauge symmetry under $\Phi_a \rightarrow \Phi_a + c$; interchange
symmetry under $\sigma_z \rightarrow -\sigma_z$, $\Phi_a \rightarrow
\Phi_a + \frac{\pi}{\sqrt{2}}$,   $\Phi_b \rightarrow
\Phi_b - \frac{\pi}{\sqrt{2}}$; and time reversal symmetry under
$x\rightarrow -x$,  $\Phi_{a,b} \rightarrow
-\Phi_{a,b}$. By then
performing a unitary transformation using
\begin{equation}
\Lambda' =
 \exp\left[ i\frac{V_1}{\sqrt{2}\pi}\Phi_a(0)+
 i\frac{V_2}{\sqrt{2}\pi}\Phi_b(0) \right]
\end{equation}
the first two terms in Eq(\ref{Uab})
vanish so that the transformed Hamiltonian becomes
\begin{equation}
\Lambda' H \Lambda'^{-1} = H_o + \frac{V_3}{\pi \tau_c} \sigma_z
\cos[\sqrt{2} \Phi_b(0)] + \Delta_o \sigma_x.
\label{Ub}
\end{equation}
Noting that the second term is nothing but the third term in
Eq(\ref{Ueo'}), one might be tempted to conclude that the dynamics of
the impurity is independent of $V_2$. However,
 a term of the form $ \sigma_x \frac{\partial\Phi_b(0)}{\partial
  x}$ will appear in Eq(\ref{Uab}) from a perturbation expansion in $\Delta_o$
and $V_2$
 which {\em cannot} be eliminated by the
canonical transformation and is {\em not}
irrelevant. Indeed, this term  breaks the {\em apparent} symmetry of
Eq(\ref{Ub}) under $\Phi_b \rightarrow -\Phi_b$ which corresponds to
$\Phi'_e \leftrightarrow \Phi'_o$ and is {\em not} an exact symmetry
in the presence of $V_2$ or similar terms. In the Appendix B we
consider a more general formulation of the two site problem which
shows how, even if the part of the Hamiltonian that is symmetric
under $\sigma_z \rightarrow -\sigma_z$ is diagonalized fully before
bosonizing, the energy dependence of scattering processes will,
nevertheless, generally lead to terms which play a similar role to
$V_2$.

We now analyze the effects of the even-odd symmetry breaking terms,
such as $V_2$, and other terms that are generated from these,
 by use of a Coulomb gas representation. As
we shall see, in certain regimes, extra {\em relevant} operators are
generated from the marginal operators that will delocalize the heavy particle.

\section{Computations}
\subsection{Path Integral Representation}
In this section, we analyze the bosonized Hamiltonian Eq(\ref{Ueo'})
by a Coulomb gas representation and show how extra operators are
generated which physically correspond to the impurity hopping together
with one or more electrons. These will delocalize the impurity in
regimes in which it was previously believed to be localized. (They will
also generate extra important operators in  Hamiltonians with ``
electron assisted tunnelling'' like that analyzed by Vlad\'{a}r
and Zawadowski\cite{Zawadowski}.) The Hamiltonian Eq(\ref{Ueo'}) is in a
convenient form since it includes, in a simple way, a term that breaks
the artificial symmetry in the absence of $V_2$, as well as a simple
form of the impurity coupling to the Fermi sea. To treat
Eq(\ref{Ueo'}) we first perform a canonical transformation from $H$ to
$H'$ using the unitary  operator
\begin{equation}
\Lambda =
 \exp\left[- i\left(\frac{V_1}{\sqrt{2}\pi}\Phi_e(0)+
 \frac{V_3}{\sqrt{2}\pi}\sigma_z \Phi_o(0)\right) \right].
\label{Lambda}
\end{equation}
Then $H$ becomes
\begin{equation}
H'=\Lambda H \Lambda^{-1}= H_o + \frac{V_2}{\pi \tau_c} \cos[\sqrt{2}
\Phi_o(0)] + \Delta_o \left\{ \sigma_+ \exp\left[i\sqrt{2} Q_o
  \Phi_o(0)\right] + h.c. \right\}
\label{Hres}
\end{equation}
where $\sigma_\pm =\frac{1}{2} \left(\sigma_x \pm i\sigma_y \right)$
and $Q_o=-\frac{V_3}{\pi}$ which
as will be seen later is the effective charge for the hopping process.
In Appendix A it will become clear why $Q_o$ really is the charge
transferred when the impurity hops between far away sites, by expressing
$Q_o$ in terms of the scattering phase shifts.

The
Hamiltonian Eq(\ref{Hres}) is expressed entirely in terms of
exponentials of boson operators which are particularly convenient for
deriving a Coulomb gas representation. Note also that the even parts
of the Bose field are completely decoupled from the odd parts and the
impurity. Although the formally irrelevant operators in Eq(\ref{Ueo'})
have been ignored, their effects under the canonical transformation
would only be to modify the coefficient of the $V_2$ term, and to give
operators that are still irrelevant and break no symmetries, although
they would include coupling to the even part of the Bose field. We can
however safely ignore these.

The correlations of the impurity position
$\sigma_z$, will not be affected by the canonical transformation as
$\sigma_z$ commutes with $\Lambda$. We therefore work with
Eq(\ref{Hres}) and drop the prime on $H$. Including the effects of spin
$\sigma$, we have, dropping the ``o'' (odd) subscript on $\Phi$
\begin{equation}
H= H_o + \frac{V_2}{\pi \tau_c}\sum_\sigma \cos[\sqrt{2}
\Phi^\sigma(0)] + \Delta_o \left\{ \sigma_+ \prod_\sigma \exp\left[i\sqrt{2}
Q_o
  \Phi^\sigma(0)\right] + h.c. \right\}
\label{Hsigma}
\end{equation}
with
\begin{equation}
 H_o= \sum_\sigma \int_{0}^{\infty} \frac{d\epsilon}{2\pi} \epsilon\,
\phi^{\sigma+}(\epsilon) \phi^{\sigma}(\epsilon) e^{-\epsilon \tau_c}
\label{Hosigma}
\end{equation}
and
\begin{equation}
\Phi^\sigma(x)= \int_{0}^{\infty}
\frac{d\epsilon}{\sqrt{2\pi\epsilon}} \left[ \phi^\sigma(\epsilon)
e^{i\epsilon x} + \phi^{\sigma+}(\epsilon) e^{-i\epsilon x} \right]
e^{-\frac{\epsilon\tau_c}{2}}.
\label{Phisigma}
\end{equation}

We are interested in the zero temperature partition function
$Z=Tr \left[e^{-\beta H} \right]$ in the limit $\beta \rightarrow \infty$.
If we expand Z in $V_2$ and $\Delta_o$ we obtain a sum over one
dimensional ``paths'' from $\tau=0$ to $\beta$. Each of these paths
corresponds to a process in which the impurity hops between the sites
at particular times shifting the phase of electron excitations, while
at other times the electrons hop via the $V_2$ term. Such a path is
illustrated in Figure 2. For simplicity, we first work with a {\em single
spin} species. Then we can write Z as
\begin{eqnarray}
Z=\sum_{n=0}^\infty  \sum_{m=0}^\infty \; \prod_{k=1}^{2m}
\left( \sum_{\zeta_k=\pm1} \right) \; \Delta_o^{2n}
 y^{2m} \delta_{0,\sum_{k=1}^{2m}\zeta_k} \nonumber \\
\int_0^\beta ds_1 \ldots \int_0^{s_{2m-1}} ds_{2m} \;
 \int_0^\beta d\tau_1 \ldots \int_0^{\tau_{2n-1}} d\tau_{2n}
Z_{nm}\left( \{ \zeta_k \},\{s_i \},\{\tau_j\} \right)
\label{Z1}
\end{eqnarray}
with
\begin{equation}
Z_{nm}\left( \{ \zeta_k \},\{s_i \},\{\tau_j\} \right) =
\left< 0 \left| T \left[
e^{i\zeta_{2m}\sqrt{2}\Phi\left(s_{2m}\right)} \ldots
e^{i\zeta_1\sqrt{2}\Phi\left(s_1\right)}
e^{-iQ_o \sqrt{2}\Phi\left(\tau_{2n}\right)} \ldots
e^{iQ_o \sqrt{2}\Phi\left(\tau_1\right)}
\right] \right| 0 \right>.
  \label{Znm}
\end{equation}
where the product over different $\zeta_k=\pm1$ corresponds to the two
different terms in
$\cos\left(\sqrt{2}\Phi\right)=\frac{1}{2}\left(e^{i\sqrt{2}\Phi}
+e^{-i\sqrt{2}\Phi}\right)$ and the ``fugacity'' $y$ of the electron
hops is defined as
\begin{equation}
y\equiv \frac{V_2}{2\pi\tau_c}
  \label{fugacity}
\end{equation}
We have taken the impurity to be on the ``one'' site at $\tau=0$,
this merely reduces Z by a multiplicative factor of two. Note that the
signs of the $\pm i \sqrt{2} Q_o \Phi\left(\tau_n \right)$ must
alternate corresponding to the particle hopping back and forth,
i.e. alternating $\sigma_+$ and $\sigma_-$ terms from
Eq(\ref{Hsigma}),  and the sum is constrained to an even number of
hops because the impurity begins and ends at the same site. In
addition,
we shall see that only terms with an even number of $y$ ``charges''
will contribute, thus the sum $\sum_{k=1}^{2m} \zeta_k$ has to vanish. We
have suppressed the dependence of $\Phi(x=0)$ on the variable $x$ and by
$\Phi(\tau)$ we denote:
\begin{equation}
\Phi(\tau) = e^{+H_o\tau} \Phi  e^{-H_o\tau}
  \label{Phitau}
\end{equation}
with the expectation in Eq(\ref{Znm})  taken with the ground state of
$H_o$. The evaluation of the expectation value of the time ordered
product in Eq(\ref{Znm}) is particularly simple due to the bosonic
character of $\Phi$. First we observe that Eq(\ref{Phitau}) becomes
\begin{equation}
\Phi(\tau)= \int_{0}^{\infty}
\frac{d\epsilon}{\sqrt{2\pi\epsilon}} \left[ \phi(\epsilon)
e^{-\epsilon \tau} + \phi^+(\epsilon) e^{\epsilon \tau} \right]
e^{-\frac{\epsilon\tau_c}{2}}.
  \label{Phitau2}
\end{equation}
Thus it is of the form $\Phi(\tau)= B(\tau)+B^+(-\tau)$ where
$B(\tau)$ is a boson having commutation relations:
\begin{eqnarray}
\left[B\left(\tau\right),B\left(\tau'\right)\right]&=&0  \nonumber \\
\mbox{and }\left[B\left(\tau\right),B^+\left(-\tau'\right)\right]&=&
\int_0^\infty \frac{d\epsilon}{\epsilon} e^{-\epsilon\tau_c}
e^{-\epsilon\left(\tau-\tau'\right)} = I(\tau-\tau')
  \label{Acommutations}
\end{eqnarray}
for $\tau > \tau'$, $I(\tau-\tau')$ being formally divergent at small energies;
but only
the finite part
\begin{equation}
\tilde{I}(\tau)\equiv I(\tau)-I(0)= \ln\left(\frac{\tau_c}{\tau+\tau_c}\right)
  \label{Itau}
\end{equation}
will enter physical quantities. By the standard procedure of
reordering the operators to bring $B$ to the right and $B^+$ to the
left using the commutators in Eq(\ref{Acommutations}), and noting that
for zero temperature, only the ground state of
$H_o$ will appear on the right and left, we see that terms with the
total charge $\sum_k \zeta_k \neq 0$ will give negative infinite terms
in the exponentials and thus zero contribution to $Z$. Furthermore the
partition function $Z_{nm}$ can be written in terms of effective
interaction between the charges with strength
\begin{equation}
{\cal W}_{ij}= 2 q_i q_j \tilde{I}\left(\left|r_i-r_j\right|\right)
  \label{intpotential}
\end{equation}
where $r_i=\tau_i$ with $q_i=+Q_o$ for the $\sigma_+$ impurity hops
$1\rightarrow2$; $r_i=\tau_i$ with $q_i=-Q_o$ for the $\sigma_-$
impurity hops $2\rightarrow1$; and $\tau_i=s_i$ with
$q_i=\zeta_i$ for the electron hops, $y$. The partition
function $Z_{nm}$ is thus simply the Boltzmann factor for the charges
interacting with the logarithmic potential Eq(\ref{intpotential}),
 i.e.
\begin{equation}
Z_{nm}\left( \{ \zeta_k \},\{s_i \},\{\tau_j\} \right)=
e^{-{\cal E}_{nm}}
\end{equation}
with
\begin{eqnarray}
{\cal E}_{nm}=- 2 Q_o^2 \sum_{\left(l,l'\right)}^{2n} \left(-1\right)^{l+l'}
\ln\left(\frac{\left|\tau_l-\tau_{l'}\right|+\tau_c}{\tau_c}\right)
\nonumber \\
-2  \sum_{\left(k,k'\right)}^{2m} \zeta_k \zeta_{k'}
\ln\left(\frac{\left|s_k-s_{k'}\right|+\tau_c}{\tau_c}\right)
\nonumber \\
+2 Q_o \sum_{k=1}^{2m} \sum_{l=1}^{2n}
\left(-1\right)^l \zeta_k
\ln\left(\frac{\left|s_k-\tau_l\right|+\tau_c}{\tau_c}\right)
  \label{Boltzmanfactor}
\end{eqnarray}
It is instructive to note that the exponential of minus the second term in
Eq(\ref{Boltzmanfactor}) has the same form as the amplitude $\left( A_m
\right)^2$
in Eq(\ref{Asubn}) which is nothing else than the square of an m
particle Greens function. The reason it is the square of $A_m$ and not just
$A_m$ is that here we have the product of the Greens functions of {\em two}
sets of particles, the $c_1^+$'s and the $c_2^+$'s.

The partition function $Z$ in Eq(\ref{Z1}) with $Z_{nm}$ from
Eq(\ref{Boltzmanfactor}), is thus a 1-D Coulomb gas with logarithmic
interactions between integer charges $\pm \zeta_k$ with fugacity $y$
and ``hopping'' charges with fugacity $\Delta_o$ and charge $\pm
Q_o$ which must strictly alternate in sign.

To take into consideration the effects of more than one spin species,
we must modify Eq(\ref{Boltzmanfactor}) to include the effective
interactions between the impurity hops $\sigma_\pm$ from each spin and
the $\pm 1$ charges from the $V_2$ term in H for each spin species.
Since the Bose fields $\Phi^\sigma$ are independent, the interactions
will be simply additive. Thus we must replace the ``charges'' $q_i$ in
Eq(\ref{Boltzmanfactor}) by vector charges $\vec{q_i}$ with two
components for spin-1/2 electrons. For the impurity hops we have
\begin{eqnarray}
\begin{array}{rcrcc}
\vec{q}&=&\vec{Q_o}&\equiv& \left(Q_o,Q_o\right) \nonumber \\
 $or$ \ \ \ \
\vec{q}&=&-\vec{Q_o}&\equiv& \left(-Q_o,-Q_o\right)
\end{array}
 \label{Qovec}
\end{eqnarray}
for $1\rightarrow2$ or $2\rightarrow1$ respectively, while for the
electron hops, we have
\begin{eqnarray}
\begin{array}{rcl}
\vec{q}&=&\left(\zeta, 0\right) \nonumber \\
$or$ \ \ \ \
 \vec{q}&=&\left(0, \zeta\right)
\end{array}
  \label{zetavec}
\end{eqnarray}
for spin $\uparrow$ or $\downarrow$ electrons respectively with
 $\zeta = \pm 1$, depending on the direction of the electron hop. The
 total effective action now has the form
\begin{eqnarray}
S_{nm}&=&{\cal E}_{nm} - 2m \ln y - 2n \ln \Delta_o \nonumber \\
&=& -2 \sum_{j>i} \vec{q_i} \cdot \vec{q_j}
\ln\left(\frac{r_j-r_i+\tau_c}{\tau_c}\right) - 2m \ln y - 2n \ln \Delta_o
   \label{Seff2}
\end{eqnarray}
with the ordered times $r_i=s_i$ or $\tau_i$  and the $2n$ charges
$\vec{q}_i=\pm
\vec{Q}_o$  with fugacity $\Delta_o$, strictly alternating in
sign. The $2m$ charges in Eq(\ref{zetavec}) with fugacity $y$, need not
alternate in sign but the sum of all their charges must be zero.

\subsection{Renormalization Group Analysis}
We are now in the position to analyze the behavior of the impurity in
terms of the properties of the generalized Coulomb gas with the action
of Eq(\ref{Seff2}). We are
particularly interested here in whether or not the impurity can be
localized. Thus we consider the effects of a small hopping rate
$\Delta_o$ and analyze the Coulomb gas perturbatively in both
$\Delta_o$ and $y$. Standard balancing of the energy and ``entropy''
terms indicate that $y$ is exactly marginal --- as it must be ---, while the
hopping $\Delta_o$ has renormalization group (RG) eigenvalue
\begin{equation}
\lambda_o= 1 - 2Q_o^2
\label{lambdao}
\end{equation}
under rescaling of the small time cutoff,
$\tau_c$, with the factor of 2 coming from the two spin species. Since
for $Q_o>1/\sqrt{2}$, $\lambda_o<0$, it would appear that the impurity
hopping is irrelevant in this case, thereby leading to the conventional
conclusion
that a charge two impurity, which has $\vec{Q}_o=\left(1, 1\right)$
corresponding to the two spin channels, can
be localized. As we shall see, however, extra charges will be
generated under renormalization which invalidate this conclusion.

Since there are two types of charges $\vec{\zeta}$ and $\pm
\vec{Q}_o$, of which only the $\pm \vec{Q}_o$ are restricted to
alternate in sign, there are various processes which can be regarded
as composite charges. For example, a charge  $\vec{q}=\left(Q_o-1,
Q_o\right)$  can be formed if an impurity hop and
an electronic hole hop are close to each other. This process as well as the
more general processes which generate charges  $\left(
Q_o \pm n, Q_o \pm n\right)$ with $n=1,2,3,\ldots$ were
discussed physically in the Introduction. It is clear that while they
do not exist in the original Hamiltonian they are generated under
renormalization (or from perturbation theory). For example the
hopping matrix element $\Delta_{\left(-1,0\right)}$ of a process with charge
$\left(
Q_o-1, Q_o\right)$ will be generated under
renormalization with magnitude proportional to $y\Delta_o$. Thus, in
general, we must consider all possible composite charges and their
effects on each other.

We denote charges associated with general types of impurity hops:
$\vec{Q}$ and $-\vec{Q}$  for $\sigma_+$ and $\sigma_-$ hops,
respectively,   which occur at times $\tau_i$; $\vec{\zeta}$  for purely
electronic hops
  at times $s_i$; and   charge $\vec{q}$  for generic
 hops of either type at times $r_i$ with fugacity $z$. We need to
analyze the effects of integrating out all pairs of charges with
spacings between $\tau_c$, the cutoff and hence the minimum allowed
spacing, and $\tau_c(1+\delta l)$  with $e^l$ the time rescaling
factor. Pairs of charges, $\vec{q}_1$ and $\vec{q}_2$, which do not
sum to charge zero, will generate composite charges
$\vec{q}_1+\vec{q}_2$ with fugacity $z_1 z_2 k_{12} \tau_c \delta l$
with the $k \tau_c \delta l$ factor from the possible separations and
ordering of $\vec{q}_1$ and $\vec{q}_2$, with $k=2$ unless both
$\vec{q}$'s are impurity hops, in which case $k=0$ if they are both
$\sigma_+$ and $k=1$ (since only one ordering is possible), if one is
$\sigma_+$ and the other $\sigma_-$. Neutral pairs, i.e. those with
total  charge zero, will
not produce composite charges but will screen the interactions between
the remaining charges.

We can proceed as usual by considering the
effects of one neutral pair on other charges, specifically on a charge
$\vec{q}$ at time $r$. If the pair is $\pm \vec{\zeta}$, i.e. purely
electronic, then the
effect of the two possible orderings cancel (up to modifying
sub-logarithmic corrections to the interaction between remaining
charges) and we thus ignore these. The interesting case is thus a pair
of hops. The allowed orderings of a $\pm
\vec{Q}$ pair at times $\tau\pm \tau_c/2$ with fugacities
$\Delta_{\vec{Q}}=\Delta_{-\vec{Q}}$, depends on $\sigma_z(\tau)$.
Thus the interaction of this pair with charge $\vec{q}$ at time $r$ is
 \begin{equation}
\delta I_{\vec{q}}(r) = -2\tau_c \vec{Q}\cdot \vec{q}
\frac{\sigma_z(\tau)}{\tau-r}
\label{delta I}
\end{equation}
for $|\tau-r|\gg  \tau_c$, the appropriate limit for analyzing the
renormalization of the long time interactions. Expanding $e^{-S}$ in
$\delta  I_{\vec{q}}(r)$ and integrating over the possible position,
$\tau$, of the pair and the intra-pair spacing in the range $\tau_c$
to $\tau_c (1+\delta l)$ we see that there are contributions every
time $\sigma_z(\tau)$ changes sign, i.e. at times $\tau_i$. This
generates an effective interaction between $\vec{q}$ and all impurity
hops, but {\em not} between $\vec{q}$ and  purely electronic hops. For an
impurity hop
$\pm \vec{Q}_i$ at time $\tau_i$, the generated  effective
interaction is
\begin{equation}
\delta I_{\vec{q},\pm\vec{Q}_i}=\pm 4 \tau_c^2 \Delta_{\vec{Q}}^2
\vec{Q}_i \cdot \vec{q} \, \ln\left|\tau_i-r\right| \; \delta l.
  \label{delta Iq}
\end{equation}
This thus has the effect of modifying the interaction of each
$\vec{Q}_i$ with {\em all} other charges by a way that is equivalent
to changing $\vec{Q}_i$ by
\begin{equation}
\delta \vec{Q}_i = -2 \tau_c^2 \Delta_{\vec{Q}}^2 \vec{Q}\, \delta l.
\label{deltaQi}
\end{equation}
with $\pm\vec{Q}$ the charges of the electron impurity impurity hop
pair that have been integrated out.
Since each $\vec{Q}_i$ is of the form $\vec{Q}_o$ plus an integer
vector, we see that the net effect is just to change $Q_o$ by
\begin{equation}
\delta \vec{Q}_o = -2\tau_c^2 \delta l \sum_{\vec{N}}
\left(\vec{Q}_o+\vec{N}\right) \Delta_{\vec{N}}^2
  \label{deltaQo}
\end{equation}
with the sum running over all possible types $\vec{Q}$ of $\sigma_+$
charges, i.e. $\vec{Q}=\vec{Q}_o + \vec{N}$ with $\vec{N}$ an integer
vector; here and henceforth we use the abbreviated notation
\begin{equation}
\Delta_{\vec{N}}
\equiv \Delta_{\vec{Q}_o + \vec{N}}.
\label{DeltaN}
\end{equation}
To this order in
$\Delta_{\vec{N}}$ and $y_{\vec{N}}$, the fugacities for multi-electron
hops which can have all integer vectors $\vec{N}$ except $ \left(
0, 0\right)$, the RG flow equations are, after absorbing
$\tau_c$'s into $\Delta$ and $y$ to make them dimensionless
\pagebreak
\begin{eqnarray}
\frac{d\vec{Q}_o}{dl}&=&-2\sum_{\vec{N}} \left(\vec{Q}_o+\vec{N}\right)
\Delta_{\vec{N}}^2 \nonumber \\
\frac{d\Delta_{\vec{N}}}{dl}&=&\left(1-\left|\vec{Q}_o+\vec{N}\right|^2\right)
\Delta_{\vec{N}}+
2\sum_{\vec{N}' \neq 0} y_{\vec{N}'}\, \Delta_{\vec{N}-\vec{N}'}
\nonumber \\
\frac{dy_{\vec{N}}}{dl}&=&\left(1-\left|\vec{N}\right|^2\right)
y_{\vec{N}} + \sum_{\vec{N}' \neq 0} y_{\vec{N}'}\,
y_{\vec{N}-\vec{N}'}+ \sum_{\vec{N}'} \Delta_{\vec{N}'}\,
\Delta_{\vec{N}-\vec{N}'}
  \label{floweqns}
\end{eqnarray}
with
\begin{equation}
\Delta_{\vec{N}}\equiv\Delta_{\vec{Q}_o+\vec{N}}=
\Delta_{-\left(\vec{Q}_o+\vec{N}\right)}
\end{equation}
by the $1\leftrightarrow2$ interchange symmetry. As can be seen  from
Eq(\ref{floweqns}) all the multi-electron
hop terms are irrelevant, thus we need only retain $y_{\left(1
  0\right)} = y_{\left(0 1\right)}\equiv y$; from Eq(\ref{floweqns})
we see that this is sufficient to generate all the composite charges
with fugacities of order $\Delta_o$ times powers of $y$.

{}From Eq(\ref{floweqns}) we see that, generically for two spin
channels, there are at least {\em three} relevant operators for any $Q_o$:
$\Delta_{\left(\left[Q_o\right]-1,\left[Q_o\right]\right)}$ and
$\Delta_{\left(\left[Q_o\right],\left[Q_o\right]-1\right)}$ with
$\left[Q_o\right]$ the fractional part of $Q_o$ are always relevant
while either $\Delta_{\left(\left[Q_o\right],\left[Q_o\right]\right)}$
or $\Delta_{\left(\left[Q_o\right]-1,\left[Q_o\right]-1\right)}$ or
both will also be relevant. More generally, we arrive at the same conclusion as
from the simple physical argument of the Introduction: in order to
localize the impurity, more than four channels (including spin) are
needed so that, if each channel, $\gamma$, is optimally coupled by a
$\frac{1}{2}$-integer $Q_\gamma$, then $\sum \,Q_\gamma^2 \, >1$ and
the impurity can be localized.

 The RG equations (Eq(\ref{floweqns})) are quite different from those in the
literature: if there
were no $V_2$ term, then the extra composite impurity hopping terms
would not be generated, and the impurity could appear to be localized
by just $s$-wave scattering. Note that  the apparent $c_1 \rightarrow
e^{i\phi}c_1$
symmetry when $V_2=0$ actually allows for charges $\vec{Q}=\vec{Q}_o +\vec{N}$
with {\em even }
integer vector $\vec{N}$, but these will not prevent localization. This
point suggests that $Q_o$ should be defined up to an even integer
(i.e. mod 2). In fact it will be seen in Appendix~A that $Q_o$ is
actually $\pi^{-1}$ times a phase shift  which naturally leads to its
consideration of  mod 2.

In the Hamiltonian considered by Vlad\'{a}r and
Zawadowski\cite{Zawadowski} single electron assisted hopping terms  that
correspond to
$\Delta_{\left(-1,0\right)}$ and $\Delta_{\left(0,-1\right)}$ are
present but their  Hamiltonian has the implicit symmetry
$c_{1}\rightarrow - c_{1}$,  $d_1 \rightarrow -d_1$. Then
 only a subset of the $\Delta_{\vec{N}}$ can be generated,
specifically those with $N_{\uparrow}+N_{\downarrow}$ {\em odd };
again this artificial extra symmetry will change the behavior by limiting the
number of possibly
relevant operators.

\section{Discussion and Conclusions}

In the previous section, we have seen that an impurity hopping between
two symmetrically placed sites cannot be localized unless it is
coupled strongly to more than four spin and angular momentum ``channels''. If
the sites are
nearby --- as they must be if the bare hopping rate is to be
appreciable --- then one cannot use angular momentum channels, and must,
instead,
generalize the treatment. One way to do this, which shows directly the
role of the irrelevant operators and relies on no symmetries other
that the site  interchange symmetry, is to use the
one-electron eigenstates of the symmetrized electron Hamiltonian $H_S$
which is the average of the Hamiltonians with the impurity on the two
sites. The antisymmetric part then scatters the electrons between even
and odd parity states of $H_S$. The analysis in this representation is
carried out in Appendix B, with the same conclusions being reached as
in Section III.

The problems with most earlier treatments of the two-site system have
been of two types: In many of the treatments, an extra
$U\left(1\right)$ symmetry associated with the independence of the
electrons which interact with the impurity at the two sites is
implicitly assumed\cite{Chang}. The $V_2$ term that breaks this symmetry is
marginal but it creates extra operators, particularly those which move
one localized hole with the impurity, and these processes delocalize a
charge two particle which had previously claimed to be localized if it
interacts with only $s$-wave electrons at each impurity.

Recently,
there has also been a substantial literature on the
relationship between an impurity with ``electron assisted tunnelling''
--- i.e. hopping of the impurity concomitantly with the motion of {\em
  one } electron of either spin --- and the two channel Kondo problem
with the $z$-component of the Kondo ``spin'' being the impurity
position and its $x$-component the hopping, (i.e. our $\sigma_z$ and
$\sigma_x$). The two ``channels'' of the Kondo problem are then the two
electron spin
species which are exactly degenerate in the absence of an external
field. The frequently used Hamiltonian\cite{Murumatsu} for this problem was
introduced by Vlad\'{a}r and Zawadowski\cite{Zawadowski}. However,
they completely neglected a $V_2$-like term. They derive the
electron-assisted hopping term assuming that $Q_o\ll 1$ and $k_F R
\ll 1$. Even in this case, using their numbers we find that the
estimate of the amplitude of the electron-assisted hopping term they
get is smaller by at least one to two orders of magnitude from the
amplitude of $\Delta_{\left(-1,0\right)}$ that is generated
from Eq(\ref{floweqns}) after renormalizing to $l=O(1)$:
\begin{equation}
\Delta_{\left(-1,0\right)} \sim y\Delta_o \sim \Delta_o \pi \rho_F V
  \label{deltaamplitude}
\end{equation}
with $k_F R \leq 1$. In the case of $k_F R \geq 1$, essential to get
$Q_o=O(1)$ which is the relevant situation for localization, their derivation
of the electron-assisted hopping term
breaks down. In our treatment and because of the existence of $V_2$ we
show how both $\Delta_{\left(-1,0\right)}$  and other relevant terms (e.g.
$\Delta_{\left(-1,-1\right)}$) naturally arise. Thus, in any case, we
believe we have here a more complete physical picture of the problem.

 As we have seen, these extra terms change the physics
for large $Q_o$ and small hopping --- the ``weak coupling'' (small J) limit in
the
Kondo language. Most of the recent
literature\cite{Affleck,Emery,Affleck2,Cox,Andrei}  has focused on the
``strong coupling'' behavior of the Kondo system i.e. the regime at low
energies with
parameters, particularly $Q_o$, such that exchange is relevant and flows to
large values corresponding to the particle hopping back and forth
between the two sites. Novel non-Fermi liquid behavior has been
found theoretically in this regime for which our considerations are
not directly relevant. Nevertheless, the extra symmetry implicit in
these treatments is potentially dangerous, indeed, as we will show
elsewhere, terms that break this symmetry change the physics in the strong
coupling delocalized regime, as they did in the weak hopping regime
analyzed in this paper.

One of the major motivations for the present work  was the hope of
gaining further understanding about the properties of a heavy particle
that can hop on sites of a periodic lattice impeded (or in some
regimes assisted!) by the coupling to a Fermi sea. This has potential
relevance for the mobility of muons in metals\cite{Kagan}, the sharpness (or
rounding) of X-ray edge singularities when the deep hole can move
(albeit with
a large bare mass), and possibly the properties of a heavy d- or
f-electron band coupled to a light conduction band\cite{Si}. The main papers
(e.g. reference 1 and references therein)  on
the behavior of a single particle in such a periodic system, suffer
from some of the same problems as those on the two-site system: they
treat a subset of the allowed operators and do not allow for the
effects of others that may be generated. Not surprisingly, the
conclusions of these papers are the same as for the two site case:
that a charge two particle with only $s$-wave scattering can be
localized while a charge one particle cannot be. In light of the
present results, this conclusion should clearly be reexamined.

In the
spirit of the work of Sols and Guinea\cite{Sols}, one could treat
each step of the particle motion --- whether via nearest or further
neighbor hopping --- essentially independently and look at the
renormalization of each such hopping term separately, including the possible
motion of electrons with the particle,  by the methods
outlined in this paper. From this approach the following conclusion
would be immediate: that the particle {\em cannot} be localized unless
all the hopping processes are irrelevant, and this can only occur if
there are more than four electron channels involved for every possible
hopping process and the coupling is sufficiently strong. Thus with just
$s$-wave scattering, a particle {\em
  cannot} be localized, in contrast to the conclusion in the
literature\cite{Sols,Yamada}. Instead, the particle will move around with a
screening cloud of
electrons in tow.

Unfortunately, there are problems in extending the
two site results in this way. The primary one is the spatial structure
of the system and the lack of independence between the electrons
involved in hopping between different pairs of sites. For any finite
number of sites, the electrons can be treated as essentially one
dimensional at sufficiently low energies, and generalization of the
present methods can be used to categorize all the operators. We
have explicitly carried out this procedure for a simple case of three
sites symmetrically arranged in a triangle and find that the same
conclusions are obtained as in the two site case. We believe that this
should likewise hold for any finite number of sites. However for an
infinite lattice of sites, the electrons must be considered to be
fully three dimensional and the limit of small bare hopping rate that
we have studied perturbatively, may not be exchangeable with the limit
of an infinite number of sites. In particular, one has to worry about at
least two effects. First, even if the particle is moving very slowly,
there will always be particle-hole excitations with group velocities
{\em slower} than the particle; these can perhaps not be treated in
the same manner as the rapidly moving excitations. Second, one could
argue that there are an infinite number of scattering ``channels''
involved because of electrons near each of the sites. Perhaps these
might make it easier to localize the particle, although it is not
clear how this could come about. Conversely, they might somehow
interfere and prevent localization even if there are many scattering
channels at each site. We must, unfortunately, leave these questions for future
study. But we should note, that if the preliminary conclusion about the
difficulty of
localizing a heavy particle is correct, it may have implications for
some of the suggested possibilities for interesting new physics with
a heavy f-electron band coupled to a light conduction electron band\cite{Si}.
Again, we
leave this, as well as the possibility of interesting effects on the
{\em electrons} caused by a delocalized impurity, for future work.

The present analysis, although not introducing new techniques, has, we
hope, made the physical picture behind the competition between
orthogonality catastrophe and hopping clearer and simpler. In
addition, the pitfalls of the standard combination of bosonization
techniques and ``large'' canonical transformations have been brought
out and should be heeded by workers on other problems in this area.

\acknowledgments
We thank Bert Halperin, Andy Millis and Igor Smolyarenko for useful
discussions. This work was supported in part by the NSF via grant DMR
9106237.

\appendix
\section{}
In this Appendix it is  illustrated how one can get a Coulomb
gas representation of the partition function $Z$ directly from the original
form of the Hamiltonian Eq(\ref{Heff}), without bosonizing. At the
same time the physical interpretation of $Q_o$ as the transferred
electron screening charge instead of simply being the interacting
potential $V_3$ --- which is the naive result of bosonization  --- will become
apparent. The
method used here has been applied to systems very similar to ours in
the past\cite{Chang,NdD,Anderson} so we will not go into the details
of the calculations.

It is straightforward to see that using the Hamiltonian in
Eq(\ref{Heff}) one can get an expression for the partition function $Z$ similar
to Eq(\ref{Z1}):
\begin{equation}
Z=\sum_{n=0}^\infty  \sum_{m=0}^\infty \;
 \Delta_o^{2n}  y^{2m}
\int_0^\beta ds_1 \ldots \int_0^{s_{2m-1}} ds_{2m} \;
 \int_0^\beta d\tau_1 \ldots \int_0^{\tau_{2n-1}} d\tau_{2n}
Z_{nm}\left( \{s_i \},\{\tau_j\} \right)
  \label{Zfermion1}
\end{equation}
where $y \propto V_2$ and
\begin{equation}
Z_{nm}\left( \{s_i \},\{\tau_j\} \right)  =
\left< 0 \left| T \left[
c^+_1(s_1) c_2(s_1) \ldots c^+_2(s_{2m}) c_1(s_{2m})
\exp\left(- \int_0^\beta d\tau' H'_o\left(\tau'\right) \right)
\right] \right| 0 \right>.
  \label{Zfermionnm}
\end{equation}
In Eq(\ref{Zfermionnm}) by $H'_o$ we denote
\begin{equation}
H'_o(\tau)=H_o + \left(V_1+V_3\right)c_1^+ c_1 + \left(V_1-V_3\right)c_2^+ c_2
  +2 V_3 \theta(\tau) \left( c_2^+ c_2 - c_1^+ c_1 \right)
  \label{Ho'}
\end{equation}
with
\begin{equation}
\theta(\tau) = \left\{ \begin{array}{lcl}
                    0 & \mbox{ \ \ \ for \ \  $\tau \epsilon
                     \left(\tau_{2k},\tau_{2k+1} \right)$} &\ \ \ \ \
k=0,1,\ldots,n \\
                    1 & \mbox{ \ \ \ for \ \  $\tau \epsilon
                     \left(\tau_{2k-1},\tau_{2k} \right)$} &\ \ \ \ \
k=1,\ldots,n
                   \end{array}
             \right.
  \label{zetatau}
\end{equation}
Since the $V_2$ term contains both $c_2^+ c_1$ and $c_1^+ c_2$ which
can be generated at any time $s_i$, there is no constraint in the order in
which they appear in Eq(\ref{Zfermionnm}). However, due to the
number-conserving character of
$H'_o(\tau)$ there is a constraint of having, for each $i$, as many $c_i^+$ as
$c_i$'s in the time ordered product of $Z_{nm}$. For convenience we pick $|0>$
to be the ground state of $H_o+
 \left(V_1+V_3\right)c_1^+ c_1 + \left(V_1-V_3\right)c_2^+ c_2$.
We can see that $Z_{nm}$ represents a
Green's function of m particles, types ``1'' and ``2'', in a time dependent
potential which changes its value
when the impurity hops at times $\tau_i$. This, in the language of the Kondo
problem\cite{Anderson}, is equivalent to a path in which the impurity
spin flips at times $\tau_i$ while the electrons flip their ``spin''
(i.e. 1--2 index)
at times $s_j$. The amplitude for such a process is a product of  two
independent parts, i.e. the amplitude of the 1-particles and that of the
2-particles. Now since the
``interaction'' term $V_3$ is really only a one particle operator,
 {\em all} the diagrams will be either closed
loops or else will  end at times $s_i$, so we can thus treat
them independently\cite{NdD}. The closed loop contribution is simply $Z_{n0}$
which is identical to that of the Kondo problem.

 In order to get the
other contribution let us first consider m=1. Then the amplitude
corresponds to  a one-particle Green's function in the presence of the time
dependent Hamiltonian which can be shown to have the following long
time (i.e. all time differences $\left|\tau-\tau'\right|\gg\tau_c$)
solution\cite{Anderson}:
\begin{equation}
G(s,s')\propto \frac{1}{s-s'}
\prod_{k=1}^n
\left|\frac{\left(\tau_{2k}-s\right)}{\left(\tau_{2k}-s'\right)}
\frac{\left(\tau_{2k-1}-s'\right)}{\left(\tau_{2k-1}-s\right)}
\right|^{\frac{\delta}{\pi}}
\label{Gtt'}
\end{equation}
with the phase shift\cite{Perakis}
\begin{equation}
\delta=-\arctan\frac{V_3}{\beta}+\pi\Theta(-\beta )
  \label{delta/pi}
\end{equation}
with $\Theta$ the Heavyside step function and
\begin{equation}
\beta=1-V_1\tan\theta+
\frac{1}{4}\left(1+\tan^2\theta\right)\left(V_1^2-V_3^2\right)
  \label{beta}
\end{equation}
with $\theta$ related to the short time behavior of $G_o$ (the
propagator of $H_o$) including the existence of bound states and particle-hole
asymmetry\cite{NdD}.
Note that the extra $\Theta$-function, which in the past had often been
ignored\cite{Chang,NdD} and was only recently
introduced explicitly\cite{Perakis} makes $\delta$ defined between
$-\frac{\pi}{2}$ and $\frac{\pi}{2}$.

{}From Eq(\ref{delta/pi}) it
becomes clear that although $V_1$ may be formally decoupled from the
impurity in the bosonized version of the problem (see Eq(\ref{Ueo'}))
it does renormalize the exponent $\delta$. This leads us to the conclusion
that indeed $\delta$ is a non-universal quantity --- not in general
simply related to the phase shifts for scattering off a static
impurity\cite{Yamada}.

More generally, since electron operators anti-commute, the amplitude for m
particles is  a determinant of $G(s_i,s'_j)$ terms\cite{Anderson} (see
Eq(\ref{Asubn})):
 $\det G\left(s_{i},s'_{j}\right)$, where $s_i$ and $s'_j$ are times at
 which the i-th  1-particle was
created and the j-th 1-particle was annihilated, respectively.
Expanding the determinant (over i,j) out, we get
\begin{equation}
 \det G\left(s_{i},s'_{j}\right)=  \frac{\prod_{i>j} \lgroup s_i-s_j
   \rgroup \prod_{i>j} \lgroup s'_i-s'_j \rgroup }
{\prod_{i,j} \lgroup s_i-s'_j \rgroup}
\prod_i \prod_{k=1}^n
\left|\frac{\left(\tau_{2k}-s_i\right)}{\left(\tau_{2k}-s'_i\right)}
\frac{\left(\tau_{2k-1}-s'_i\right)}{\left(\tau_{2k-1}-s_i\right)}
\right|^{\frac{\delta}{\pi}}
\label{Gss'}
\end{equation}
We can now bring this in a form closer to Eq(\ref{Boltzmanfactor}) if we
introduce a number $\zeta_i$ which is +1 for $s_i$ and -1 for $s'_i$.
Then, quoting the result for the closed loop amplitude\cite{Chang,Anderson}
\begin{equation}
 \exp\left[\left(\frac{\delta}{\pi}\right)^2 \sum_{l>l'}
 \left(-1\right)^{l+l'} \ln\left|\tau_l-\tau_{l'}\right| \right],
  \label{closedloops}
\end{equation}
 and using Eq(\ref{Gss'}), after taking the square of all the above amplitudes
 to take into account the two  channels (1,2), $Z_{nm}$ becomes:
\begin{eqnarray}
Z_{nm}\left( \{ \zeta_k \},\{s_i \},\{\tau_j\} \right)=
&\,&\exp\left[ +2\left(\frac{\delta}{\pi}\right)^2
\sum_{\left(l,l'\right)}^{2n} \left(-1\right)^{l+l'}
\ln\left|\tau_l-\tau_{l'}\right|\right]
\nonumber \\
\times &\,&\exp\left[ +2  \sum_{\left(k,k'\right)}^{2m} \zeta_k \zeta_{k'}
\ln\left|s_k-s_{k'}\right|\right]
\nonumber \\
\times &\,&\exp\left[-2 \frac{\delta}{\pi}  \sum_{k=1}^{2m} \sum_{l=1}^{2n}
\left(-1\right)^l \zeta_k
\ln\left|s_k-\tau_l\right|\right]
 \label{Znmdelta}
\end{eqnarray}
This is equivalent to Eq(\ref{Boltzmanfactor}) if $\frac{\delta}{\pi}
\rightarrow Q_o$. The essential
reason for the appearance of an arctangent of the potential in
Eq(\ref{delta/pi}) is because  the
singular effects of higher order  terms in $V_3$ are included.
Eq(\ref{delta/pi})
makes $\delta$ finite even if $V_3$ is infinite. Finally, from
Friedel's sum rule,  it would seem that $\frac{\delta}{\pi}$ corresponds to the
electronic screening charge moved when the impurity hops from site to site,
thus $Q_o$ would be just this screening charge.
One should emphasize, however that this correct only in the limit of
large site separation for which $V_2$ is small and $\delta$ is not
appreciably renormalized from the phase shift off a static impurity at
one of the two sites.

In conclusion, we have seen here that $|Q_o| \leq \frac{1}{2}$, since
$-\frac{\pi}{2} < \delta \leq \frac{\pi}{2}$. However this is  essentially
equivalent with the results of this paper in which we have  shown that for
any initial bare value of $Q_o$ the most relevant process that will
dominate in delocalising the impurity, is the one that has charge q
with absolute value
\begin{equation}
\left|q\right| =\min_{n\epsilon Z}\left\{
\left|\left[Q_o\right]\right|,  \left|\left[Q_o\right]-n\right|\right\}
\leq \frac{1}{2}
\end{equation}
with $[Q_o]$ the fractional part of $Q_o$ (see discussion after
Eq(\ref{floweqns})).  The physical picture presented in this paper
provides an intuitive way of understanding why the branch of $\delta$
implied by Eq(\ref{delta/pi}) is dominant. This was not apparent in some of
the earlier work\cite{Sols,Yamada,Hamann,Chang,Schotte,NdD} and is the
source of some of the erroneous claims  about particle
localization\cite{Sols,Yamada}.

\section{}
In this Appendix, we show how the site problem can be analyzed generally in
terms of ``channels'' even in the absence of any symmetry except
the equivalence of the two sites. In addition, we will see how
bosonization in a representation in which no $V_2$-like electron
hopping term can exist
will still, if handled carefully, yield terms which play the same
role. Furthermore, the possibility of exchange of ``charge'' between
various channels will be found, yielding another, albeit related way,
that charges can be reduced, and localization impeded. We consider the
electronic Hamiltonians
with the impurity at either site one or two, denoted $H_1$ and $H_2$, and
diagonalize exactly the {\em symmetrized} Hamiltonian
\begin{equation}
H_S=\frac{1}{2} \left(H_1+H_2\right).
  \label{Hsymmetric}
\end{equation}
The full Hamiltonian is then written as
\begin{equation}
H=H_S+\sigma_z H_A +\Delta_o \sigma_x
  \label{Hfull}
\end{equation}
with the antisymmetric part
\begin{equation}
H_A=\frac{1}{2}  \left(H_1-H_2\right).
  \label{Hantisym}
\end{equation}
With fermion operators that diagonalize $H_S$, we see that neither
$V_2$- nor $V_1$-like terms can occur. At each energy, there will be a
countable degenerate set of scattering eigenfunctions of $H_S$, which we label
by $k=\epsilon-\epsilon_F$, setting $v_F=1$; and for the states that
are  even (e)  under interchange $1\leftrightarrow 2$ we use a ``channel''
index $\eta$ which we
can choose later for convenience, while we use an index $\theta$ for the odd
(o) states. Thus
\begin{equation}
H_S=\sum_k \sum_\eta k c^+_{k\eta e} c_{k\eta e}+\sum_k \sum_\theta k
c^+_{k\theta o} c_{k\theta o}
  \label{Hsym2}
\end{equation}
and the antisymmetric part has the general form
\begin{equation}
H_A=\sum_{k,k'} \Gamma_{\eta,\theta}(k,k')\left[ c^+_{k\eta e} c_{k'\theta
  o} + h.c.\right].
  \label{Hanti2}
\end{equation}
Since we are interested in the behavior near the Fermi surface, we can
choose linear combinations of the even states at $k=0$ and likewise
the $k=0$ odd states, so that
\begin{equation}
\Gamma_{\eta,\theta}(0,0)=\delta_{\eta\theta} \gamma_\eta
  \label{Gammaetatheta}
\end{equation}
is diagonal in $\eta, \theta$; then in this representation we can denote
$\theta$ also by $\eta$. We now form
the operators
\begin{equation}
\Psi_{\eta e(o)}= \frac{1}{\sqrt{N_{e\left(o\right)}}} \sum_k
 c_{\eta k e\left(o\right)}
  \label{Psietae}
\end{equation}
with appropriate normalization factors $N_{e(o)}$ as in Eq(\ref{Neo}),
and define \begin{eqnarray}
\Psi_{\eta 1} = \frac{1}{\sqrt{2}} \left( \Psi_{\eta e} +\Psi_{\eta
  o} \right) \nonumber \\
\Psi_{\eta 2} = \frac{1}{\sqrt{2}} \left( \Psi_{\eta e} -\Psi_{\eta
  o} \right).
\label{Psieta12}
\end{eqnarray}
The Hamiltonian then takes the form
\begin{equation}
H=H_S + \sigma_z \sum_\eta \gamma_\eta \left( \Psi^+_{\eta 1}
\Psi_{\eta 1} -\Psi^+_{\eta 2} \Psi_{\eta 2}\right) + \sigma_z {\cal M}
  \label{Hfull2}
\end{equation}
where the correction term $\cal M$ involves the deviations of
$\Gamma$ from the diagonal form for $k,k' \neq 0$:
\begin{equation}
{\cal M} = \sum_{k,k'} \sum_{\eta,\eta'}
\left(\Gamma_{\eta \eta'}(k,k')-\gamma_\eta \delta_{\eta \eta'}
\right) \left( c^+_{k\eta e}c_{k'\eta' o} + h.c. \right).
  \label{varepsilon}
\end{equation}
If we bosonize the $\Psi_{\eta 1,2}$ following the prescription used
in Section II.B, then in the absence of $\cal M$, the
Hamiltonian just takes the simple form
\begin{equation}
H= \Delta_o \sigma_x + \sum_\eta H_\eta
  \label{Hsimple}
\end{equation}
with
\begin{equation}
H_\eta= K_\eta + \frac{\gamma_\eta}{\sqrt{2}\pi} \sigma_z
\frac{\partial\Phi_{\eta o}\left(0\right)}{\partial x}
  \label{Heta}
\end{equation}
where $K_\eta$ is the kinetic energy of the $\eta$ bosons. We then have
for each channel $\eta$ an independent charge
\begin{equation}
Q_{\eta o}=\frac{\gamma_{\eta o}}{\pi},
\end{equation}
so that\cite{footnote2}
\begin{equation}
\alpha_o=\sum_\eta Q_{\eta o}^2.
\label{alphao2}
\end{equation}
In the absence of $\cal M$, the particle would thus be
localized if $\alpha_o>1$.

Analysis of the form of $\cal M$, shows, with the $\eta$ basis
chosen to vary slowly with $k$, the existence of formally irrelevant
terms like
\begin{equation}
\sigma_z \frac{\partial \Phi_{\eta o}(0)}{\partial x}
\cos\left[\sqrt{2}\Phi_{\eta o}\left(0\right) \right],
  \label{irrevterm1}
\end{equation}
the $\frac{\partial}{\partial x}$ essentially arising from terms in
$\cal M$ linear in $k$ and $k'$.
Under the canonical transformation that eliminates the $\sigma_z \frac{\partial
\Phi_{\eta o}(0)}{\partial x}$
terms in Eq(\ref{Heta}), these will generate
$\cos\left[\sqrt{2}\Phi_{\eta o}\left(0\right) \right]$
terms which are of exactly the same form as these that would have
arisen from a $V_2$ term originally. These terms create integer charges
which can then combine with the $Q_\eta$ charges to give effective
charges of $Q_\eta-n_\eta$ with integer $n_\eta$. Because of the
choice of the $\eta$ fermions, these terms no longer have quite the
interpretation of moving $n_\eta$ electrons with the impurity.
Physically, this is quite simple: in the
correct basis, the electrons do not need to be moved, they will do so
on their own due to the change in the electronic part of H as the
impurity moves.

In addition to $V_2$-like terms, terms of the
form
\begin{equation}
e^{i\frac{1}{\sqrt{2}}\left(\Phi_{\eta e}-\Phi_{\eta' e}\right)}
\sin\left(\frac{\Phi_{\eta o}}{\sqrt{2}}\right)
\,\sin\left(\frac{\Phi_{\eta' o}}{\sqrt{2}}\right)
  \label{halfchargemove}
\end{equation}
will also be generated under the canonical transformation from
$\Gamma_{\eta \eta'}$ terms linear in $k$. These will create
$\frac{1}{2}$ charges in the formerly-decoupled even channels. For any
pair $\eta$, $\eta'$, it can be seen that the effective charge
squared of combining this process with a hop has a contribution to
$\alpha_o$ from
these two pairs of even and odd channels of
\begin{equation}
\left(Q_\eta-\frac{1}{2}\right)^2 +\frac{1}{4} +
\left(Q_{\eta'}-\frac{1}{2}\right)^2  +\frac{1}{4}.
  \label{halfchargesqred}
\end{equation}
Since the resulting contribution from each even-channel charge is
$\frac{1}{4}$, it can be seen  that the effective charge squared of
each channel cannot, in the general case, be reduced below $\frac{1}{4}$,
i.e. the same result as in the absence of the channel mixing terms.
Therefore these terms do not change the conclusion of the earlier discussion.
Nevertheless, the presence of channel mixing terms will complicate the
analysis of the many site problem and could perhaps change the physics
in a spatially extended system.




\begin{references}
\bibitem{footnote1} \mbox{$\alpha, \beta,
\theta$ must satisfy $\alpha^2+\beta^2=\frac{\pi}{k^2}\left[1-
\left( \frac{\sin{kR}}{kR} \right)^2 \right]^{-1}$ and} \\
\mbox{$2\alpha\beta\cos{\theta}= -\frac{\pi}{k^2} \frac{\sin{kR}}{kR}
\left[1- \left(\frac{\sin{kR}}{kR} \right)^2 \right]^{-1}$.}
\bibitem{footnote2}Note that $\alpha_o$  defined in
  Eq(\ref{A2a}) had an extra factor of 1/2 multiplying the sum. The
  reason this should not appear here is that there we were
  essentially overcounting, by a factor of two, the number of channels by
treating the 1-
  and 2-electrons separately. It is clear that  this factor of two
  should {\em not} appear in Eq(\ref{alphao2}) as $Q_{\eta o}$ is
  defined as the charge transfer of the {\em combined} process
  of both 1- and 2-electrons.
\end{references}
\end{document}